# Integrating Commercial and Social Determinants of Health: A Unified Ontology for Non-Clinical Determinants of Health

Navya Martin Kollapally, MS[1], Vipina Kuttichi Keloth, PhD[2], Julia Xu, PhD[3], James Geller, PhD[1]
[1]New Jersey Institute of Technology, Newark, New Jersey, USA; [2]Yale University, New Haven, Connecticut, USA; [3]JX consulting, Texas, USA

**Abstract**
*The pivotal impact of Social Determinants of Health (SDoH) on people's health and well-being has been widely recognized and investigated. However, the effect of Commercial Determinants of Health (CDoH) is only now garnering increased attention. Developing an ontology for CDoH can offer a systematic approach to identifying and categorizing the commercial factors affecting health. Those factors, including the production, distribution, and marketing of goods and services, may exert a substantial influence on health outcomes. The objectives of this research are 1) to develop an ontology for CDoH by utilizing PubMed articles and ChatGPT; 2) to foster ontology reuse by integrating CDoH with an existing SDoH ontology into a unified structure; 3) to devise an overarching conception for all non-clinical determinants of health and to create an initial ontology, called N-CODH, for them; 4) and to validate the degree of correspondence between concepts provided by ChatGPT with the existing SDoH ontology.*

**Introduction**
According to World Health Organization (WHO), the social determinants of health refer to the conditions that influence people's well-being, including their birth, upbringing, living and working environment, and access to healthcare[1]. These conditions are impacted by broader factors such as economics, social policies, politics, and commercial factors that affect health. *Commercial determinants of health* are situations, actions and omissions of business entities that affect individual and population health[2]. These determinants, driven by activities in pursuit of profit, include factors such as access to healthy food options, marketing and advertising strategies, as well as workplace practices. For example, marketing and advertising strategies used by corporations can impact consumer behavior and choices, potentially leading to unhealthy behaviors and lifestyles. Consequently, these factors can impact modifiable risk behaviors such as tobacco use, unhealthy diet, lack of physical activity, and harmful alcohol consumption, leading to overweight and obesity, elevated blood pressure, increased blood glucose levels, high cholesterol, and ultimately life-threatening diseases such as heart disease, cancer, liver cirrhosis, chronic respiratory disease, and diabetes. These non-communicable diseases may include lifestyle diseases and mental health issues. When non-communicable diseases are not under the control of individuals, but instead are caused by commercial activities, they could be called "industrial epidemics" or corporate-driven diseases[3]. Cardiovascular diseases account for most deaths among non-communicable diseases (17.9 million people annually), followed by cancers (9.3 million), chronic respiratory diseases (4.1 million), and diabetes (2.0 million). It is estimated that in the United States 88% of deaths annually are caused by such ailments, as well as 14% of premature deaths (dying at an age between 30 to 70)[4].

In the literature, various definitions and frameworks have been proposed to describe CDoH. Kickbusch et al.[2] define CDoH as "private sector strategies and approaches for promoting products and choices detrimental to health." The authors identify consumer and health behavior, individualization, and choices as subcategories of CDoH at the micro level, while at the macro level they include "global risk society," the "global consumer society," and the "political economy of globalization." West et al.[5] describe CDoH as "factors influencing health stemming from the profit motive." Drawing on existing CDoH and SDoH definitions, Lacy-Vawdon et al.[6] define CDoH as "a series of systems that initially materialize around systems of commercial and/or corporate power." While acknowledging that CDoH systems' influences may be positive or negative, the primary focus must be on preventing and reducing harm according to the author. Mialon[7] developed a framework for CDoH based on Kickbusch et al., which lists as factors 1) the production of unhealthy commodities by corporations; 2) the use of business, market and political practices that are harmful to health; and 3) global drivers of ill-health, shaped by the practices of corporations. Nevertheless, the author notes that there is limited research on CDoH, a lack of attention to the global drivers of ill-health, and limited studies on the activities of industries other than the food, alcohol, and tobacco industries, that contribute to CDoH. Overall, these definitions and frameworks highlight the complex interplay between commercial activities and health outcomes, and the need for further research and interventions to address CDoH and promote health equity.

As alluded to above, the existing body of literature has extensively studied the impact of commercial and corporate interests on population health. However, integrating these factors within the CDoH framework is a nascent area of research[3,5,6]. The current definitions of CDoH fail to address the linkage between CDoH and risk behaviors associated with non-communicable diseases[3]. Additionally, these definitions do not take into consideration both the positive and negative impact of CDoH factors on population health. Hence, there is a need to standardize the concepts and categorization of CDoH to cover this terminology gap. To this end, one of our objective is to develop an ontology for CDoH to address these issues. According to Gruber[8], an ontology is a formal and explicit specification of a shared conceptualization of a desired domain of interest. Ontologies have the potential to combine diverse information sources on the schema level and can be leveraged for information retrieval from unstructured text by elevating keywords to the level of ontological concepts and relationships. By developing an ontology for CDoH, we aim to provide a means of standardizing knowledge management efforts under a common conceptual model within this specific domain[9].

In order to develop an ontology, it is essential to have a comprehensive list of terms/concepts that cover the domain under consideration[10]. To enrich a domain ontology, the developers often rely on relevant research articles to gather concepts extending the depth and breadth of the ontology. In this study, we used PubMed Central (PMC) as a source for identifying relevant research in this field and eventually for harvesting concepts. However, even with extensive search, it can be challenging to gather all relevant concepts and ensure that the ontology is comprehensive. NLP techniques and chatbots have been used for the task of text summarization. GPT (Generative Pretrained Transformer)[11] models are a type of language model, which has been trained on large datasets of text. ChatGPT[12] is a chatbot developed by the company OpenAI. It is built on top of OpenAI's GPT-3.5/4 family of large language models, and it is fine-tuned using both supervised and reinforcement learning techniques. In this research, we supplement our search strategy, utilizing ChatGPT, that can generate human-like responses to natural language prompts. We propose a novel human-AI collaborative concept collection approach for developing an ontology for CDoH utilizing ChatGPT to expand our concept set.

To comprehensively integrate the influences of commercialization within the CDoH framework, it may be necessary to broaden the scope of the existing paradigms to address ways in which they impact the SDoH. For example, business policies and practices related to employment and working conditions can impact the social and economic status of workers, which can have downstream effects on health outcomes. By broadening the scope of the CDoH framework to consider the ways in which commercial activities interact with SDoH, we can gain a better understanding of how these factors impact health outcomes. Hence, in this research, we are drawing on the idea of Non-Clinical Determinants of Health[13] (for which we introduce the N-CODH ontology: ***Non-Clinical Ontology of Determinants of Health***; pronounce as *en-code*), integrating the health impact of SDoH and CDoH. We released the initial version of N-CODH[14] by integrating our existing SDoH ontology (SOHO; see Glossary of all ontologies used at the end of the paper) with the CDoH ontology presented in this paper. We hypothesize that the N-CODH ontology has the potential to transform how we approach research, policy, and public health practice, providing a more comprehensive and nuanced understanding of the complex and interrelated factors that shape health outcomes.

In summary, we are developing an ontology for CDoH by utilizing PMC articles and ChatGPT. We are supporting ontology reuse by integrating CDoH with our existing SDoH ontology (SOHO). We are presenting an overarching conception for all non-clinical determinants of health, and we created an initial ontology called N-CODH. Finally, we are validating the degree of correspondence between concepts provided by ChatGPT with the SOHO ontology.

**Methods**
*Literature Review and CDoH Concept Extraction*
We developed the CDoH ontology using ontology development principles as per Noy[10]. One of the ontology design principles is content reuse from existing ontologies[15]. As such our first step was to search the NCBO BioPortal[16], which is a unified collection of various ontologies and terminologies, currently containing 1,052 of them, with 15,644,567 classes, and 36,286 properties. Ontologies in BioPortal reuse content from other ontologies to facilitate the modeling of new classes, cover a subject domain, save development work, and support applications. The domain of our ontology is the definition of health effects of CDoH, but we could not locate any such ontology in BioPortal[17]. To arrive at this determination, we performed keyword searches using: "Commercial determinants of health," "Corporate determinants of health," "Commercial drivers of ill health," "Commercial determinants of ill health," "Commercial drivers of non-communicable disease," "Commercial determinants of non-communicable devices,"

"Commercial determinants of obesity" and their variations in the *find an ontology* and *class search* fields in BioPortal. Since our searches did not yield any results, we proceeded to develop the CDoH ontology from scratch.

We utilized the Preferred Reporting Items for Systematic reviews and Meta-Analyses framework (PRISMA 2020)[18] as outlined in Figure 1. For collecting the relevant articles for developing the CDoH ontology, we did a scoping search in PubMed Central (PMC)[19] using the query: *(commercial [All Fields] AND determinants [All Fields] AND ("health"[MeSH Terms] OR "health"[All Fields]) AND +framework [All Fields]) AND ("2018/01/17"[PDat] : "2023/01/15"[PDat])*. The search returned a total of 23,342 full-text articles. After removing embargoed articles, 23,094 full-text documents were moved to the next phase of screening. In this phase, 23,071 articles were eliminated that met the exclusion criteria: a "study on subpopulation without broader implication" and those articles that "did not discuss the health/climatic impacts of CDoH in the title/abstract." We identified 23 full-text articles that did not meet the exclusion criterion. We performed forward learning (extracting relevant articles from bibliographies of identified sources) and backward learning (extracting documents that cited the identified articles). Forward learning helped us identify nonacademic articles, including policy documents and population statistics from government websites that resulted in the addition of 14 articles from outside of PMC.

One issue that ontology builders routinely confront is that they need to work with "expensive" subject matter experts and ontology experts. Ideally, contributors to an ontology should possess both subject matter and ontology expertise. To address the issue that such experts are hard to recruit, we performed a pilot study to explore the use of ChatGPT as a "contributor." We extracted unique impacts of CDoH on public health by interrogating ChatGPT. Example prompts were "impact of CDoH on health outcome," "subcategories of the health impact of CDoH," "factors that impact health due to commercial drivers and corporates," "climatic hazards from CDoH," "10 effects of climate change that cause ill-health contributed by corporates," "list 20 subcategories of factors in private sector that cause lifestyle diseases," etc. We posed several semantically similar questions and were able to extract 40 unique impacts from ChatGPT. Each of these impacts was validated by searching for corresponding articles in PMC, using the extracted impacts as our search keywords. This analysis resulted in adding 72 articles that were excluded from the previous review. After the inclusion phase we had 109 full text research articles/reports and policy documents for concept extraction. We did a manual review of these 109 documents to extract all the concepts for developing CDoH ontology. (In the future, we will revisit this step using late breaking NLP methods.)

*Development of the CDoH Ontology*
After conducting a thorough analysis of all the concepts extracted during the concept collection phase, we divided the CDoH concepts into five main categories. They are 1) elements attributed by commercial factors, 2) elements attributed by economic factors, 3) elements attributed by environmental factors, 4) elements attributed by individual factors and 5) elements attributed by social factors. We used Protégé 5.5.0[20] for implementing the CDoH ontology in Web Ontology Language (OWL). Protégé refers to "concepts" as "classes," and allows adding properties and relationships between the classes. The class "Thing" is predefined in Protégé, and is used as the root of every ontology created with it. Protégé enables users to edit ontologies in OWL and to use a reasoner to validate the consistency and coherence of the developed ontologies. We have performed consistency checking in Protégé by utilizing HermiT reasoner version 1.4.3.456 [21]. We have also added object and data properties to concepts, allowing us to capture complex relationships between elements attributed to different factors. Examples of the object properties are "*have_education_level,*" which associates "person" with "education level," and "*have_contaminants,*" which relates "available source of drinking water" with chemicals such as "radon," "fluoride," etc.

*Creating the N-CODH Ontology by integrating the CDoH ontology with three other ontologies*
Considering the overlap between commercial and social determinants, as commercial activities can influence social factors and vice versa, to effectively address this complex interplay, it is important to take a comprehensive and integrated approach. Now that we have developed the CDoH ontology, we describe our approach to developing the N-CODH ontology. We imported three existing ontologies with factors affecting nonclinical outcomes to improve the coverage and flexibility of the CDoH ontology [22]. We integrated the designed ontology with our previously developed Social determinants Of Health Ontology (SOHO[a]) available in BioPortal[23]. SOHO lays great emphasis on covering the healthcare consequences of health inequities in hospitals and among practitioners by reusing the HOME ontology[24]

---

[a1]Due to a history of confusions between SDoH and SOHO we are using the Agency FB font for the word SOHO. To further reduce confusion, we did not assign a name to the CDoH ontology developed in this paper. See Glossary at the end.

(Healthcare Ontology for Minority Equity). HOME specifically deals with healthcare impacts due to implicit bias within and outside of healthcare. By standardizing the contributors of non-communicable diseases in an ontology, we can address the challenge of heterogeneity embedded in their definitions, categorizations, and applications. Therefore, by adding SDoH concepts to the CDoH ontology, we have created an ontology that has a comprehensive coverage of non-clinical determinants of health. The development of N-CODH is a major achievement of this study. Additionally, to represent the time progression of events, we also imported the Time Event Ontology (TEO) from BioPortal into N-CODH. Data properties, such as "parts_per_million," were added to N-CODH to represent, for example, the maximum chemical contaminant levels in drinking water. We annotated N-CODH with CURIES IDs, which ensures interoperability and makes it easier to use it as gold standard for NLP tasks.

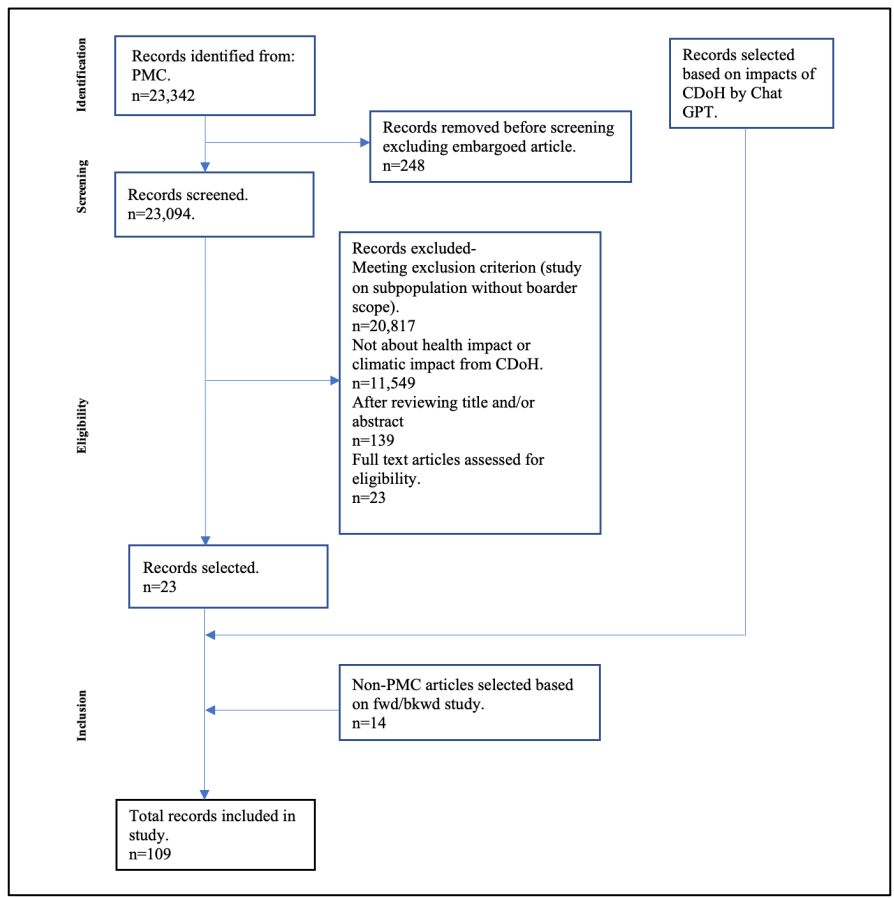

**Figure 1**. PRISMA diagram of study inclusion.

*Ontology evaluation*
Ontology evaluation is defined as the process of deciding the quality of an ontology considering a set of evaluation criteria. The four main methods of ontology evaluation are gold-standard comparison, application-based evaluation, data sources comparison, and human-centric evaluation[25]. We performed application-based and human-centric evaluation, as gold-standard and data source comparisons do not apply due to unavailability of such data to us. In addition to using the HermiT reasoner, we also used OntoMetrics[26] for application-based evaluation. Due to the absence of an existing ontology that deals with factors contributing to ill health from non-clinical determinants, we opted to add a human-centric evaluation. We involved two subject matter experts (VK, JX) with extensive experience in biomedical ontology evaluation to assess the N-CODH ontology. Below we describe this evaluation in detail.

*Application based evaluation*: The HermiT reasoner can be used to determine whether the ontology is consistent and coherent. On the other hand, OntoMetrics [26] is intended to evaluate certain aspects of ontologies and their potential for knowledge representation. Metrics provided by OntoMetrics describe domain-independent aspects of the ontology and provide deeper insights than HermiT. The OWL file developed using Protégé was uploaded to OntoMetrics as an

XML file to calculate the metrics, especially schema metrics. *Schema metrics* are used to evaluate the depth, width, richness, and inheritance of the designed ontology. Relationship richness reflects the diversity of relations and placement of relations in the ontology. Attribute richness reflects the number of attributes that are defined for each class. It can indicate both the quality of ontology design and the amount of information pertaining to instance data. Inheritance richness is a measure that describes the distribution of information across different levels of the ontology's inheritance tree or the fan-out of parent classes. This is a good indication of how well knowledge is grouped into different categories and subcategories in the ontology. Class richness is related to how instances are distributed across classes.

*Human expert evaluation*: After validating the N-CODH ontology for consistency, coherence, and semantic correctness, we utilized human expert evaluation to investigate whether the developed ontology covers the pertinent aspects of the domain under consideration correctly. We designed a spreadsheet with concept pairs of the form "Parent ← IS-A- Child" to minimize ambiguity. The parent and child concepts are connected using an IS-A relationship. Table 1 shows a snippet from the evaluation sheet with concept pairs.

**Table 1.** A snippet of the spreadsheet with concept pairs provided for evaluation to the human expert.

| Parent | Relation (same) | Child | Child? | Farther away | Reason if unrelated |
|---|---|---|---|---|---|
| Effect of climatic changes | ←IS-A- | Marketing of unhealthy food products | No | | Child concept relates to promotion of unhealthy food products and has no bearing on parent concept which relates to climate change. |
| Eating related psychopathology | ←IS-A- | Binge eating disorder | Yes | | |
| Chemical risk in drinking water | ←IS-A- | Social media affected health outcomes | No | | Health outcomes affected by social media cannot be a child of chemical risk in drinking water |
| Trade and globalisation effect on health disparities | ←IS-A- | Violating labour standards | | Yes | The concepts share a grandparent-child relationship |

Both human evaluators (VK and JX) were provided with the same spreadsheet of 100 concept pairs. Among the 100 pairs, we provided 10 concept pairs as training samples to present the flavour of the ontology and 90 pairs that needed to be evaluated. The spreadsheet contained three kinds of concepts pairs: pairs related as parent-child, pairs related as ancestor_or_grandparent-child, and pairs that were not hierarchically related. Both VK and JX were aware of the fact that the spreadsheet contained these different kinds of concept pairs. The spreadsheet contained three empty columns with the headings "Child," "Farther away," and "Reason if unrelated." The 10 samples provided to the evaluators included five of the "Child" fields filled with "No" and corresponding reasons were provided in "Reason if unrelated." Three of the "Child" fields were filled with "Yes," and two of the "Farther away" fields were filled with "Yes."

For each pair, the fourth column ("Child?" in Table 1) had to be filled with "Yes," if the evaluator felt that the concepts were connected by a parent-child (IS-A) relationship, and "No" otherwise. If the answer was "No," they were asked to fill in the reason in the column "Reason if unrelated." During the evaluation phase, these reasons provided us with directions on how to make improvements to the design of the ontology. The evaluators were asked to fill in the "Farther away" column with "Yes," whenever they felt that the concepts were related by a grandparent or ancestor relationship, i.e., a *chain* of IS-A relationships. The evaluators were also asked to give reasons in this case (in the "Reason if unrelated column"). VK and JX independently reviewed the pairs, and we used Cohen's kappa[27] test to identify the level of agreement. Cohen's kappa (κ) is a statistical coefficient that represents the degree of agreement between two raters. A κ > 0.4 is considered as moderate agreement and κ = 1 means perfect agreement. To evaluate the statistical significance of their individual results, we used Fisher's exact test[28].

*Evaluating the Concordance of the ontology with ChatGPT*

To explore the concordance of the ontology with ChatGPT, we employed the evaluation sheet developed for SDHO in previous work. The Social Determinants of Health Ontology (SDHO) is available in BioPortal[23] and had been evaluated by two medical ontology experts and a physician. For this study, ChatGPT was given concept pairs using the natural user query pattern: *"Neighborhood and built-in environment"* ←IS-A— *"Proximity to industrial facilities"* along with the question "is this a valid IS-A relationship?" The arrow was part of the input to ChatGPT. ChatGPT responded either with a positive answer (along the lines of "Yes, this is a valid parent-child relationship") or a negative answer ("No, these concepts do not share a strict IS-A relationship") along with explanations for either case. In cases where ChatGPT responded negatively, we asked follow-up questions to determine how the relationship could be defined or how the child concept could be modified. Table 2 illustrates a few of the concept pairs presented to ChatGPT. For the SDHO validation study, out of 60 concept pairs, 20 pairs shared a parent-child relationship, 20 pairs were unrelated, and the remaining 20 pairs shared a grandparent relationship (i.e., the concepts were related but not directly related).

**Table 2.** Sample of concept pairs given to ChatGPT.

| Parent | Relation | Child |
| --- | --- | --- |
| Impact of food insecurity | ←is--a | Metabolic disturbances from poor nutrition |
| Poor Housing | ←is--a | Bullying at school |
| Economic instability | ←is--a | Inability to enroll in federal assistance |
| Poor Workplace condition | ←is--a | Poor pairing of team members at work |

For those concept pairs that ChatGPT did not consider as related by an IS-A link, but instead considered it to be related by a grandparent-child relation, we experimented with a novel way of evaluation, performed by prompting ChatGPT with a series of questions diagrammatically explained in Figure 2. Subfigure a) shows that we proposed to ChatGPT that B is a child of A. However, ChatGPT indicated that it "thinks" of B as a grandchild of A. Subfigure b) represents this graphically. We then challenged ChatGPT to tell us the children of A (Subfigure c)). Interestingly, in some cases it returned B as a child of A (Subfigure d)) while in other cases it did not.

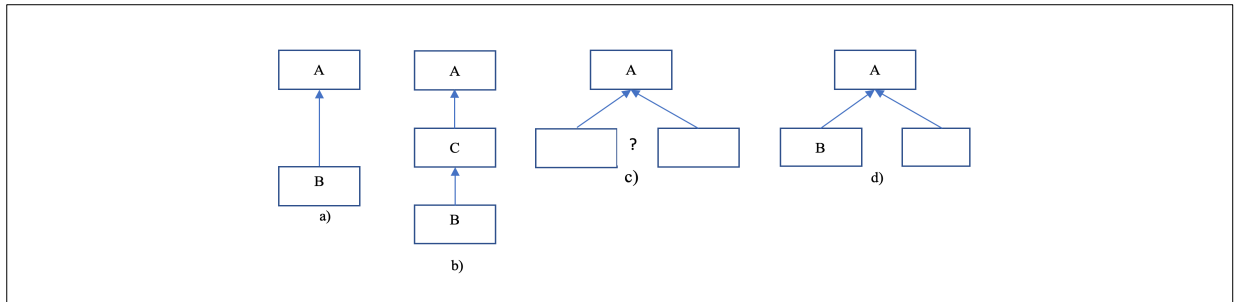

**Figure 2.** Evaluation framework for concept pairs not connected with IS-A relationship as per ChatGPT. a) Is concept B a sub concept of A? b) ChatGPT states Concept B is a grandchild of concept A. c) ChatGPT is prompted to list all the child concepts of Concept A. d) ChatGPT lists all child concepts of A *including* B, contradicting itself.

As per SDHO "Poor housing" ←is—a "pest infested house" but ChatGPT disagreed with the relation stating that: *"Poor housing" and "pest infested house" can have a distant hierarchical relationship*. With respect to ChatGPT, poor housing can encompass a variety of conditions that make a dwelling substandard, and one of those conditions could be pest infestation. Next, we prompted ChatGPT to return 10 concepts that have IS-A relationships to "Poor housing." The response from ChatGPT included *insect or pest infestation* along with other concepts such as overcrowding in house, lack of basic amenities, exposure to environmental hazards, lack of ventilation, homelessness etc. In the Results section, we will present the breakdown of these cases. A total of 276 prompts were used to get evaluation results for the 60 pairs from ChatGPT.

**Results**

*Developed N-CODH Ontology*

The CDoH ontology developed using Protégé contains 317 classes and 675 axioms along with 27 object properties and 19 data properties. Figure 3 represents the main categories and the direct subclasses of the CDoH ontology in Protégé. The IS-A relationships are indicated by indentation in the figure. N-CODH is a domain ontology that integrates the CDoH ontology with the existing SDoH ontology SOHO, the Healthcare equity ontology (HOME) and the Time event ontology (TEO). N-CODH contains 611 classes and 2603 axioms. To reference biomedical entities, Compact Uniform Resource Identifiers (CURIEs) have been added to the ontology[29]. We defined 41 object properties and 28 data properties in the first version of N-CODH. The top-level classes of N-CODH are depicted in a partial conceptual framework shown in Figure 4. The N-CODH OWL file is available on GitHub[22] and the NCBO BioPortal[14].

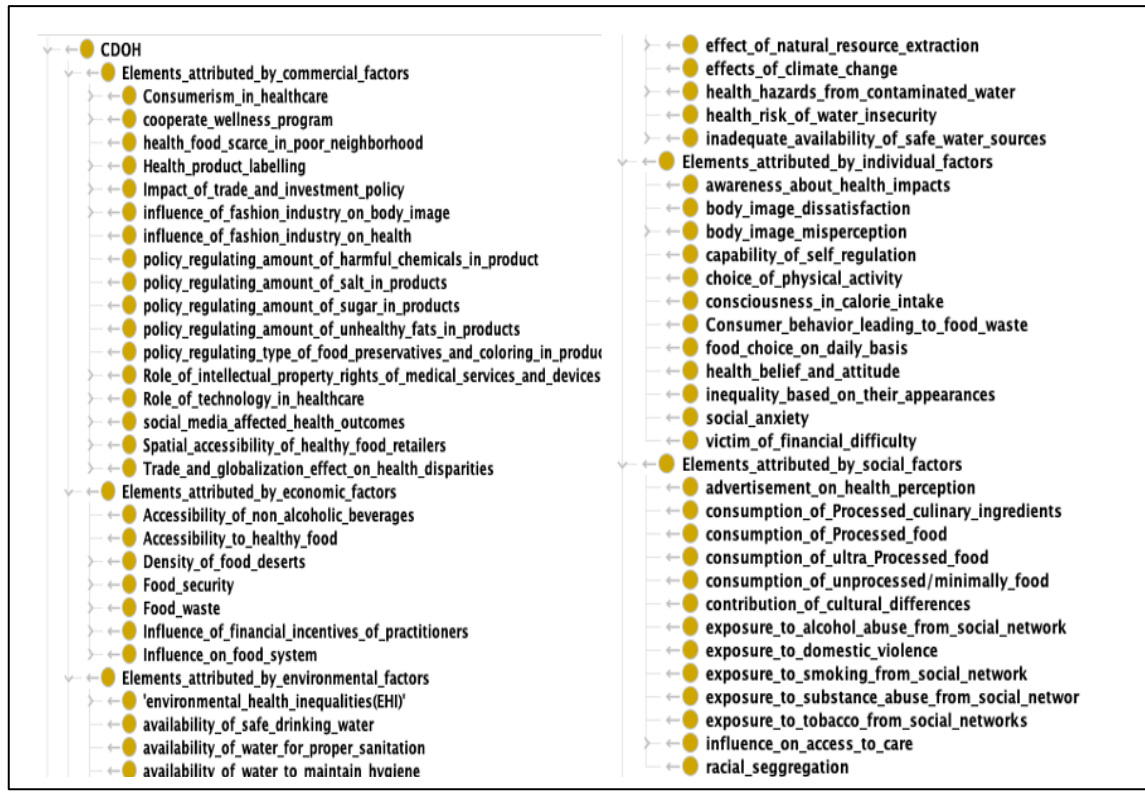

**Figure 3**. Main classes and direct subclasses of the CDoH ontology in Protégé.

*Metrics quality of N-CODH*
According to the HermiT reasoner running in Protégé, N-CODH is a coherent and consistent ontology. We performed an analysis of the ontology using OntoMetrics to obtain the schema metrics which is presented in Table 3. The N-CODH ontology aims to be a comprehensive representation covering the impacts of commercial determinants of health **and** of social determinants of health. It is characterized by low attribute richness and higher inheritance richness. The inheritance richness represents the horizontal nature of the ontology, indicating fewer levels of inheritance and a higher number of subclasses per class. N-CODH consists mainly of class-subclass relationships (as opposed to semantic relationships), leading to lower semantic relationship richness, which represents the diversity of relations and their placement in the ontology.

*Human evaluation results of N-CODH*
An evaluation by human experts was performed to ensure that the domain knowledge represented in N-CODH correctly reflects human intuitions. Both VK and JX independently evaluated 90 random concept pairs, which included 32 IS-A pairs, 14 grandparent-child pairs, and 44 unrelated pairs connected erroneously with IS-A relations. Table 4 shows the input values used to calculate Cohen's kappa. We obtained a κ=0.50502, which indicates 74.44% agreement and in turn shows that there is a fair agreement about the ontology between the two evaluators. The confusion matrix for the Fisher exact test corresponding to each evaluator is provided in Table 5 and Table 6. In the metric input, hierarchically related concept pairs include both IS-A relationships and ancestor_grandchild relationships. For both the evaluators, we obtained a p value <0.0001, which is less than p=0.05. This implies that the evaluation is statistically significant[28]. Based on the feedbacks from the experts, we renamed two of the parent concepts in N-CODH for better clarity. "Access to farmers market" was changed to "transportation access to farmers market." "Fear of deportation" was changed to "fear of deportation of illegal workers in hazardous jobs."

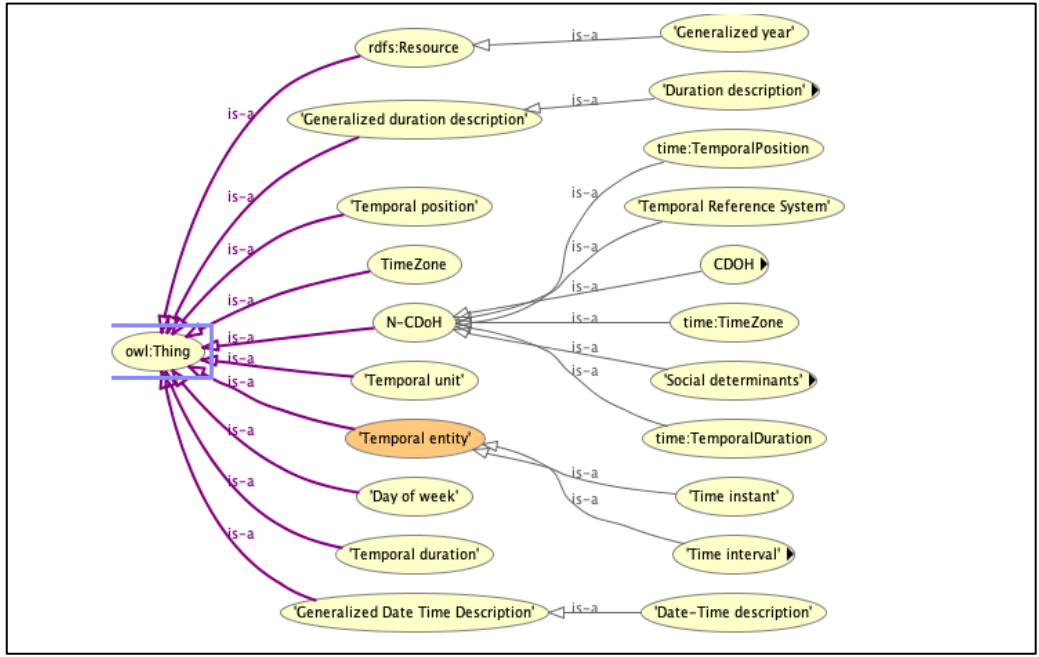

**Figure 4.** Top level hierarchical classes of N-CODH (in the figure called N-CDoH).

**Table 3.** Schema metric returned by OntoMetrics.

| Metrics | Value |
|---|---|
| Attribute richness | 0.008876 |
| Inheritance richness | 0.98816 |
| Relationship richness | 0.12336 |
| Axioms/Class ratio | 4.49905 |
| Class/relation ratio | 0.88713 |

**Table 4.** Cohen Kappa input metrics

| Description | Count |
|---|---|
| Both evaluators agree to include | 31 |
| Both evaluators agree to exclude | 36 |
| First evaluator wants to include | 3 |
| Second evaluator wants to include | 20 |

*Validation study results of ChatGPT:* ChatGPT agreed that the 20 nonrelated concept pairs taken from SOHO should not be connected by an IS-A relationship. It also correctly identified the 20 grandparent relationships. Results for the parent-child relationships were less strong. For parent-child pairs, the initial number of agreements=9, and the number of disagreements was 11. We attempted to establish the parent-child relationship for 7 of the 11 according to Figure 2.c). For 5 of the 11 pairs, children were recognized as such in the second step, corresponding to Figure 2.d). For the remaining two that was not the case. Among the remaining 4 (=11−5−2) concept pairs, 3 concept pairs were linked by "part-of" relationships and one concept pair was connected by a "type-of" relationship, according to ChatGPT. We consider the type-of relationship sufficiently similar to the parent-child (IS-A) relationship for our purposes.

**Table 5.** Confusion matrix of evaluator 1.

| Confusion matrix | Hierarchical related concept pairs | Unrelated concept pairs |
|---|---|---|
| Evaluated as hierarchical related concept pairs | 39 | 0 |
| Evaluated as unrelated concept pairs | 7 | 44 |

**Table 6.** Confusion matrix of evaluator 2.

| Confusion matrix | Hierarchical related concept pairs | Unrelated concept pairs |
|---|---|---|
| Evaluated as Hierarchical related concept pairs | 42 | 11 |
| Evaluated as unrelated concept pairs | 4 | 33 |

## Conclusions

In this project, we created an ontology to address the health impacts of CDoH, including concepts such as health hazards from climatic changes triggered by commercial actions. Using Protégé 5.5.0, we developed the CDoH ontology with 675 axioms and 317 classes along with 27 object properties and 19 data properties. Our research on CDoH indicated a need to integrate it with our previously developed Social Determinants of Health Ontology (SOHO), the Health care Ontology for Minority Equity (HOME), and the Time Event Ontology (TEO), resulting in the development of the N-CODH. The initial N-CODH ontology includes 611 classes and 2603 axioms.

To evaluate the N-CODH ontology, we utilized the HermiT reasoner and the OntoMetrics tool along with two human experts' evaluation for domain coverage. We also conducted a validation study to determine whether ChatGPT could be used to support the development of an ontology. By leveraging ChatGPT as a "contributor," we were able to supplement our publication and concept collection efforts and expand the breadth of our ontology's coverage. This human-AI collaborative approach has the potential to reduce the cost and time required to build an ontology, while still maintaining a high level of accuracy and rigor. During the validation study, ChatGPT provides us with the insight that 11 pairs out of 60 concept pairs were recognized as not strictly IS-A related. *Thus, it would be beneficial for ontology developers in general to revisit and review their parent-child pairs with ChatGPT and make necessary adjustments.* In other words, ChatGPT can be utilized as an important tool to validate additional relevant concept pairs, enriching the ontology to the desired level of granularity.

## Limitation and Future Work

One limitation of this study is that concepts were identified by human review. A second limitation is that only research articles from PMC were included. A third limitation is that the exclusion criteria that we applied could have resulted in omitting pertinent concepts. To address these gaps, in the future, NLP techniques will be utilized to extract relevant concepts from policy documents, population surveys, mortality surveys, clinical notes, scientific publications, etc. Exclusion criteria will be relaxed. As a final notable limitation, human review was limited to two experts. A third expert will be added to the team, when available.

## Glossary

SOHO: Social Determinants of Health Ontology: On BioPortal, previously developed by this team.
CDoH Ontology: Commercial Determinants of Health Ontology (No Acronym): On BioPortal, developed in this paper
HOME: Health Ontology for Minority Equity: On BioPortal, previously developed by this team.
N-CODH (*en-code*):
Non-Clinical Ontology of Determinants of Health: On BioPortal/GitHub, developed in this paper
TEO: Time Event Ontology: On BioPortal. Imported into N-CODH.

## References

1. WHO. Commercial determinants of health. [Online].; 2021 [cited 2023 February 10. Available from: https://www.who.int/health-topics/commercial-determinants-of-health#tab=tab_1.
2. Kickbusch I, Allen L and Franz C, "The commercial determinants of health.," *Lancet Glob Health,* vol. 4, no. e, pp. 895-896, 2016.
3. Lee K and Freudenberg N, "Addressing the commercial determinants of health begins with clearer definition and measurement.," *Global Health Promotion,* vol. 27, no. 2, pp. 3-5, Jun 2020.
4. WorldHealthOrganization(WHO). Noncommunicable Diseases Progress Monitor. ; 2022.
5. West R and Marteau T, "Commentary on Casswell (2013): the commercial determinants of health," *Addiction,* vol. 108:686, 2013.
6. Lacy-Vawdon Cd, B.Vandenberg , Livingstone CH. "Recognising the elephant in the room: the commercial determinants of health." BMJ Global Health. 2022; 7.
7. Mialon M, "An overview of the commercial determinants of health," *Global Health,* vol. 16, no. 74, 2020.
8. Gruber TR. Toward Principles for the Design of Ontologies Used for Knowledge Sharing. International Journal Human-Computer Studies. 1993: p. 907-928.

9. Stephan G, Pascal H and Andreas A, "Knowledge Representation and Ontologies, Berlin: Semantic Web Services," 2007.
10. McGuinness D and Noy NF, "Ontology development 101: A guide to creating your first ontology.," 2001.
11. Radford A, Wu J, Child R, Luan D, Amodei D & Sutskever I, "Language Models are Unsupervised Multitask Learners.," *Comp Sci,* 2019.
12. OPEN AI, "ChatGpt," [Online]. Available: https://openai.com/blog/chatgpt/. [Accessed Feb 2023].
13. Golembiewski E, Allen KS, Blackmon AM, Hinrichs RJ, Vest JR. "Combining Nonclinical Determinants of Health and Clinical Data for Research and Evaluation: Rapid Review." JMIR Public Health Surveill. 2019. Oct 7;5(4):e12846. doi: 10.2196/12846. PMID: 31593550; PMCID: PMC6803891.
14. Kollapally NM, Geller J. NCBO Bioportal. [Online].; 2023. Available from: https://bioportal.bioontology.org/ontologies/N-CDOH?p=summary.
15. Ochs C, Perl Y, Geller J, Arabandi S, Tudorache T and Musen MA, "An empirical analysis of ontology reuse in BioPortal," *J Biomed Inform,* vol. 71, no. 165-177, Jul 2017.
16. Whetzel P, Noy NF, Shah N, Alexander P, Nyulas C, Tudorache T and Musen MA, "BioPortal: enhanced functionality via new Web services from the National Center for Biomedical Ontology to access and use ontologies in software applications.," *Nucleic Acids Res,* pp. W541-5, Jul 2011.
17. Noy NF, Shah N, Whetzel PL, Dai B, Dorf M, Griffith N, Jonquet C, Rubin D, Storey MA, Chute CG and Musen MA, "BioPortal: ontologies and integrated data resources at the click of a mouse," *Nucleic Acids Res,* vol. 37, pp. W170-3, Jul 2009.
18. BMJ. The PRISMA 2020 statement: an updated guideline for reporting systematic reviews. ; 2021. Report No.: n71.
19. NLM. PubMed Central (PMC). [Online]. [cited 2023 Feb. Available from: https://www.ncbi.nlm.nih.gov/pmc/.
20. Musen MA. The Protégé project: A look back and a look forward. AI Matters. Association of Computing Machinery Specific Interest Group in Artificial Intelligence. 2015 Jun; 1.
21. Glimm B, Horrocks I, Motik B, Stoilos G, Wang Z. HermiT: An OWL 2 Reasoner. J. Autom. Reason. 2014; 15(3): 245-269.
22. Kollapally NM, Github. [Online]. Available from: https://github.com/navya777/N-CDoH.
23. Kollapally NM, Chen Y, Xu J and Geller J, "An Ontology for the Social Determinants of Health Domain.," IEEE -BIBM 2022 pp. 2403-2410.
24. Kollapally NM, Chen Y and Geller J, "Health Ontology for Minority Equity (HOME)," *KEOD,* pp. 17-27, 2021.
25. Raad J and Cruz C, "A Survey on Ontology Evaluation Methods.," Proceedings of the 7th International Joint Conference on Knowledge Discovery, Knowledge Engineering and Knowledge Management, p. 10.5220/0005591001790186., 2015.
26. Tello, Lozano A and Pérez AG, "ONTOMETRIC: A Method to Choose the Appropriate Ontology," *J. Database Manag.,* vol. 15, pp. 1-18, 2004.
27. Idostatistics. Cohen's kappa calculator. [Online]. [cited 2023 Feb. Available from: https://idostatistics.com/cohen-kappa-free-calculator/.
28. Fisher E test [Online].; 2023. Available from: https://www.socscistatistics.com/tests/fisher/default2.aspx.
29. language Be. Identifiers. [Online]. [cited 2023 Jan. Available from: https://biological-expression-language.github.io/identifiers/.
30. Rostock Uo. Ontometrics wiki. [Online]. Available from: https://ontometrics.informatik.uni-rostock.de/wiki/index.php/Schema_Metrics.